# Aging Time dependent Static Friction between Soft and Hard Solid Interfaces

Vinay A. Juvekar* and Arun K. Singh**,
*Retired Professor, Indian Institute of Technology Bombay, Mumbai- 400076, India
**Associate Professor, Visvesvaraya National Institute of Technology, Nagpur-440010, India
E-mail: vaj@iitb.ac.in and aksinghb@gmail.com

**Abstract**

Understanding of friction between sliding surfaces is critical for variety of applications. We present a friction model between soft and hard solid interfaces for studying aging time dependent static friction. The model is based on strengthening of dangling chains with the substrate during aging period. The friction model is, in turn, validated with the experimental data from literature. Friction properties are also estimated in terms of gelatin concentration to justify the results.

**Key Words:** Aging time friction; Gelatin gel; Population balance model; Soft solids.

## 1.Introduction

Friction occurs at all material scales ranging from atomistic to geological scales, yet many basic mechanisms of friction is not fully understood. For instance, friction experiments have established that static friction varies as logarithm of aging time as well as logarithm of shear velocity on soft solid surfaces for instance, rubbers, elastomers, hydrogels etc (Dieterich,1978; Berthoud et al., 1999; Baumberger et al., 2003; Li et al., 2011; Persson, 2013; Petrova et al., 2020). Similar observation is also seen in case of hard solid surfaces such as rocks, metals etc (Dieterich, 1979; Berthoud et al., 1999; Persson, 2013). Static friction is generally defined as the minimum force required to onset sliding of a solid from rest (Persson, 2013). It is to be noted that aging time is also known that time of stationary contact, waiting time, relaxation time etc. History dependence of static friction is critical for understanding stick-slip motion as well (Juvekar and Singh,2024). Micromechanism of aging time dependent peak friction is relatively less understood than shear rate dependent static friction (Baumberger et al., 2003;Li et al., 2011; Ouyang and Urbakh,2022). While the rate and state friction model is often used for elucidating the logarithmic behaviour of static friction, it is basically an empirical relation derived from experimental observations (Ruina, 1983). A hidden state variable is proposed to explain the effect of aging time on static friction (Ruina, 1983; Persson, 2013). Slide-hold-slide (SHS) friction test is often carried out for measuring static friction of sliding surfaces as function of aging time as well as shear rate (Dieterich, 1978; Persson, 2013; Gupta and Singh, 2016). Motivated by these observations, we propose a friction model for static friction as function of aging time and shear velocity. The model is based on previous studies concerning friction of soft solids (Singh and Juvekar, 2011; Singh et al., 2021; Juvekar and Singh, 2024)

## 2. Model Development

We divide the total time into two phases, the 'aging phase' and the 'pulling phase'. During the aging phase, the pendent chains from the gel block bond with the adsorption sites on the stationary hard surface. The process involves a number of steps include displacement of initially adsorbed water from the bonding sites as well as the adsorbing sites, diffusion of chain segments on the hard surface, displacement of short or weakly bonded chains by long and/or strongly bonded chains. We assume that most of these steps occur very rapidly and that after a very short time the number of attached chains attain a constant value. We also assume that this time is much shorter than the total aging time. During the later aging process, the number of the attached chains are assumed to remain constant. We denote the number of attached chains per unit area of contact by $N_{c0}$ . This number is decides by the nature of the gel and the nature of the hard surface and does not depend on aging time. During the rest of the aging period, conformational rearrangement of chains occur, which causes increase in the number of bonded sites, and more importantly increase in the bond strength.

During the pulling phase, the top face of the block is pulled by the puller in the detection parallel to the hard surface. We consider a case where the top surface is pulled at a constant velocity$V_0$ . The surface of the block in contact with the hard surface remains stationary until it is detached from the hard surface. If $V_0$ is the pulling velocity and $h$ is the height of the block, then the strain rate is $V_0/h$ and the shear strain over time $t$ would be $V_0 t/h$ . If $G_d$ is shear modulus of the block, the resulting shear stress is $G_d V_0 t/h$ . For the sake of simplicity we assume that this stress is equally shared by all chains. As the pulling time progresses the stress increases. This causes the chains to detach from the surface. If $N_c(t)$ is the number of the attached chains per unit area of the surface at time $t$ , counted from the beginning of the pulling phase, then the pulling force per chain is

$$f_c(t) = \frac{G_d V_0 t}{h N_c(t)} \tag{1}$$

Rate of detachment of a chain $r_b(t)$ , i.e. the fraction of the chains detached from the surface per unit time is written as

$$r_b(t) = \frac{1}{\tau} e^{-[E+W_a(t)-\lambda f_c(t)]/kT} \quad (2)$$

Here, $E$, is the thermal activation energy associated with the thermal detachment of a chain and $\tau_0$ is the characteristic time for thermal detachment, $\lambda f_c(t)$ is the strain energy stored in the chain, $\lambda$ is the activation length needed to detach a chain from the hard surface (Singh and Juvekar, 2001). $W_a$ (t) represents the adhesion energy of a chain. Its effect is to reduce the rate of breakage. Adhesion energy is expected to increase with of the aging time $t_a$. We express the time rate of change of the adhesion energy by the following equation

$$\frac{d[W_a(t_a)\,/kT]}{dt_a} = \frac{1}{\tau_a(1+t_a/\tau_b\,)} \quad (3)$$

In this equation, $\tau_a$, the characteristic fast aging time and $\tau_b$ is a slow aging time. Basis of this expression is to indicate that at the beginning of aging, adhesion energy increases faster. But as the aging time increases, the rate of aging slows down due to reduction in the degrees . This slowing down is incorporated using the term $t_a/\tau_b$ in the denominator. Solution of Eq 3 is

$$\frac{W_a(t_a)-W_0}{kT} = \frac{\tau_b}{\tau_a}\ln(1 + t_a/\tau_b) \quad (4)$$

We note that aging of the chains occurs both during waiting and pulling. Hence: $t_a = t + t_w$

$$r_b(t) = \frac{u_o}{\tau}\exp\left\{\frac{\lambda G_d V_0 t}{hkTN_c(t)} - \frac{\tau_b}{\tau_a}\ln[1 + (t_w + t)/\tau_b\,]\right\} \quad (5)$$

Where $u_o = \exp\,[-(E + W_o)/kT]$

If $t_w = 0$ and $t \ll \tau_b$ , then Eq 5 reduces to

$$r_b(t) = \frac{u_o}{\tau}\exp\left\{\frac{\lambda G_d V_0 t}{hkTN_c(t)} - \frac{t}{\tau_a}\right\} \quad (6)$$

This is the same equation, which we had used in our previous paper (Juvekar and Singh, 2024). The final equation is

$$\frac{dN(t)}{dt} = -\frac{u_o}{\tau}\exp\left\{\frac{\lambda G_d V_0 t}{hkTN_c(t)} - \frac{\tau_b}{\tau_a}\ln[1 + (t_w + t)/\tau_b\,]\right\}N(t) \quad (7)$$

When waiting time is large compared to the pulling time, then Eq 7 transforms to

$$\frac{dN(t)}{dt} = -\frac{u_o}{\tau}\exp\left[\frac{\lambda G_d V_0 t}{hkTN_c(t)} - \frac{\tau_b}{\tau_a}\ln\left(\frac{t_w}{\tau_b}\right)\right]N(t) \quad (8)$$

We are interested in time at which the last chain of the block breaks from the hard surface. We see from Eq 8 that as long as the first term of the exponent is smaller than the second, no breakage of the chain happens because the difference $\frac{\lambda G_d V_0 t}{hkTN_c(0)} - \frac{\tau_b}{\tau_a}\ln\left(\frac{t_w}{\tau_b}\right)$ is negative and hence the exponential terms is small. However, once the second term exceeds the first term, exponential becomes positive and $N(t)$ begins to decrease, Once $N_c(t)$, the first term in the exponent increases. Thus we expect rapid fall in $N(t)$ thereafter. Hence we consider that the point of slip is approximately determined by the point where the two terms in the exponent of Eq 8 become equal or the slip time is approximately given by

$$t_s = \frac{\tau_b hkTN_c(0)}{\tau_a \lambda G_d V_0}\ln\left(\frac{t_w}{\tau_b}\right) \tag{9}$$

This shows that $t_s$ is proportional to the logarithm of the waiting time and inversely propositional to pulling velocity. When $t_w$ and $t_s$ are comparable, the more accurate solution is

$$t_s = \frac{\tau_b hkTN_c(0)}{\tau_a \lambda G_d V_0}\ln\left(\frac{t_w + t_s}{\tau_b}\right) \tag{10}$$

This is an approximate analysis. Eq 7 could be solved numerically to find $t_s$ and compare it with numerical simulations.

**3.0Results & Discussion:**

Fig.1 presents static friction stress vs. time of gelatin hydrogel with c=5% (wt./vol.) plot for varying waiting time $t_w$= 5,12 & 72s (Baumberger et al., 2003). As noted earlier that peak value of friction is basically static friction of the gel block as in Fig.1. Aiming to fit dynamic friction too, magnitude of jump velocity at peak friction is also estimated from the numerical simulations (Juvekar and Singh, 2024). It is observed that jump velocity is 0.6, 0.63 & 1.08 mms$^{-1}$ for wating time $t_w$=5,12 & 72s respectively. A physical significance of jump velocity is that rupture of the sliding interface begins at higher slip velocity than the pulling velocity (Juvekar and Singh, 2024). It is also seen in that Fig.1 experimental dynamic friction fits well the dynamic friction model with aging (Singh et al., 2021; Juvekar and Singh, 2024).

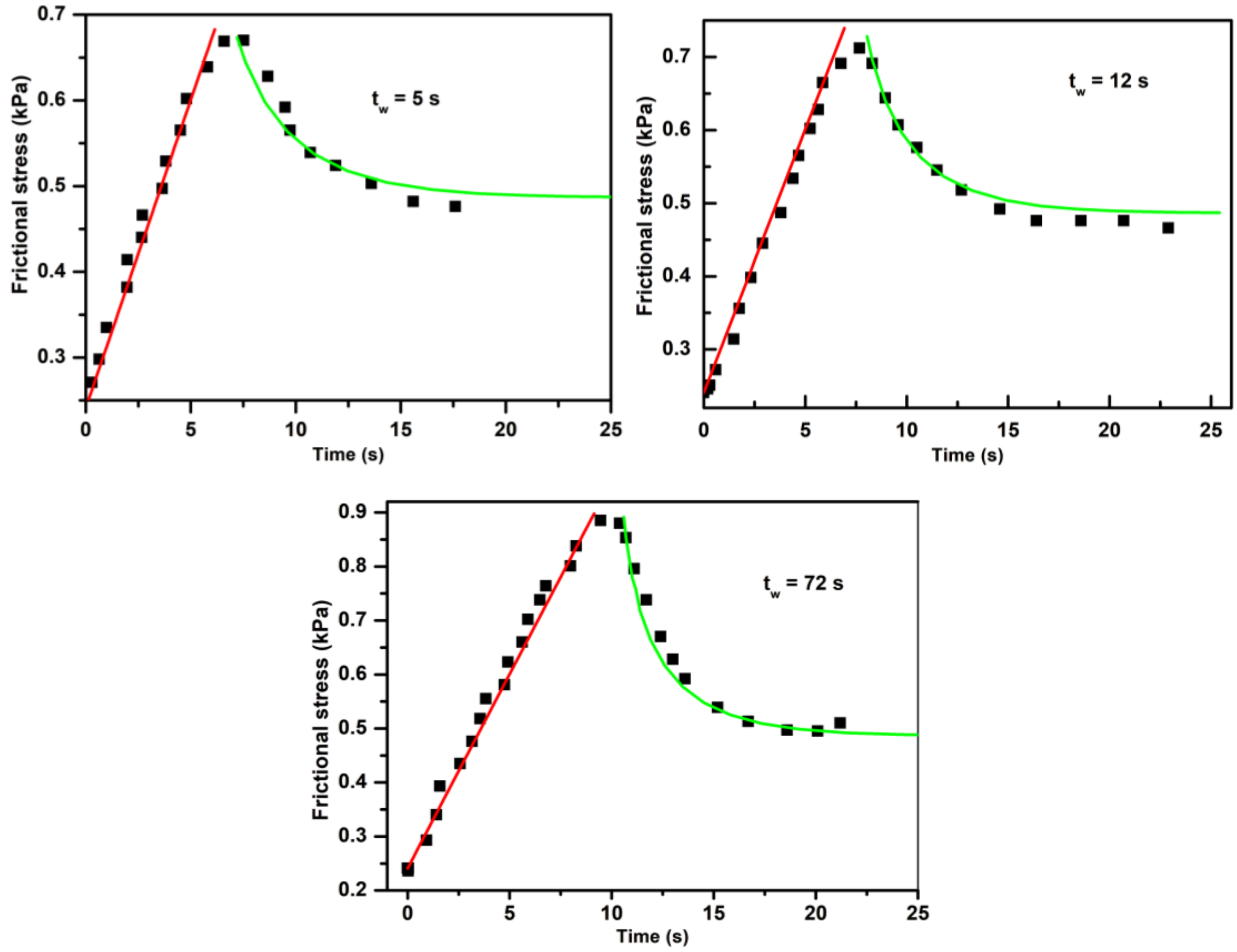

Fig.1 Frictional stress vs. time for varying waiting time $t_w$ = 5,12 & 72s of gelatin hydrogel with conc. c= 5% (wt./vol.). Experimental data were adopted from Baumberger et al. (2003).

The value of the parameters and properties is obtained in above fitting during stick period of static friction as following: $N_0 = 2.0\times10^{12}$, $\tau_a$ =20 s, $u_0 = 10^{-6}$, stiffness of the gel block (c=5% wt./vol.) $K_b$= 330.0 kPa.m$^{-1}$ at pulling velocity $V_0$= 0.30 mms$^{-1}$. While slip period: $N_0 = 1.5\times10^{14} m^{-2}$, $M = 5.5\times10^{-5}$ N.m$^{-1}$, $M = 0.5\times10^{-3} N..m^{-1}$, $\lambda$ =10$^{-10}$m, , $u_w$ =1 and $\Delta\hat{l}_m$ =5, $\alpha$ =5 , $\beta$ =0.5 and stiffness of the gel block $K_b$ =250 kPa.m$^{-1}$ at fixed pulling velocity $V_0$= 0.30 mms$^{-1}$. Noting that $V_0$ well is above to critical velocity $V_{cr}$= 0.120 mms$^{-1}$ of the gel block (Baumberger et al., 2003). The reason for difference of $N_0$ between static and dynamic friction is attributed to increased number of available sites after the rupture of interface (Juvekar and Singh, 2024).

Now using Eq.4, we have predicted the value of experimental static friction vs. aging time the gelatin gel with concentrations c=5%,8% &10% (wt./vol.) respectively (Baumberger et al., 2003). Fig.2 presents the match between the friction model and the experimental data. Numeric value of the parameters and properties related to the friction model is listed in Table 1.

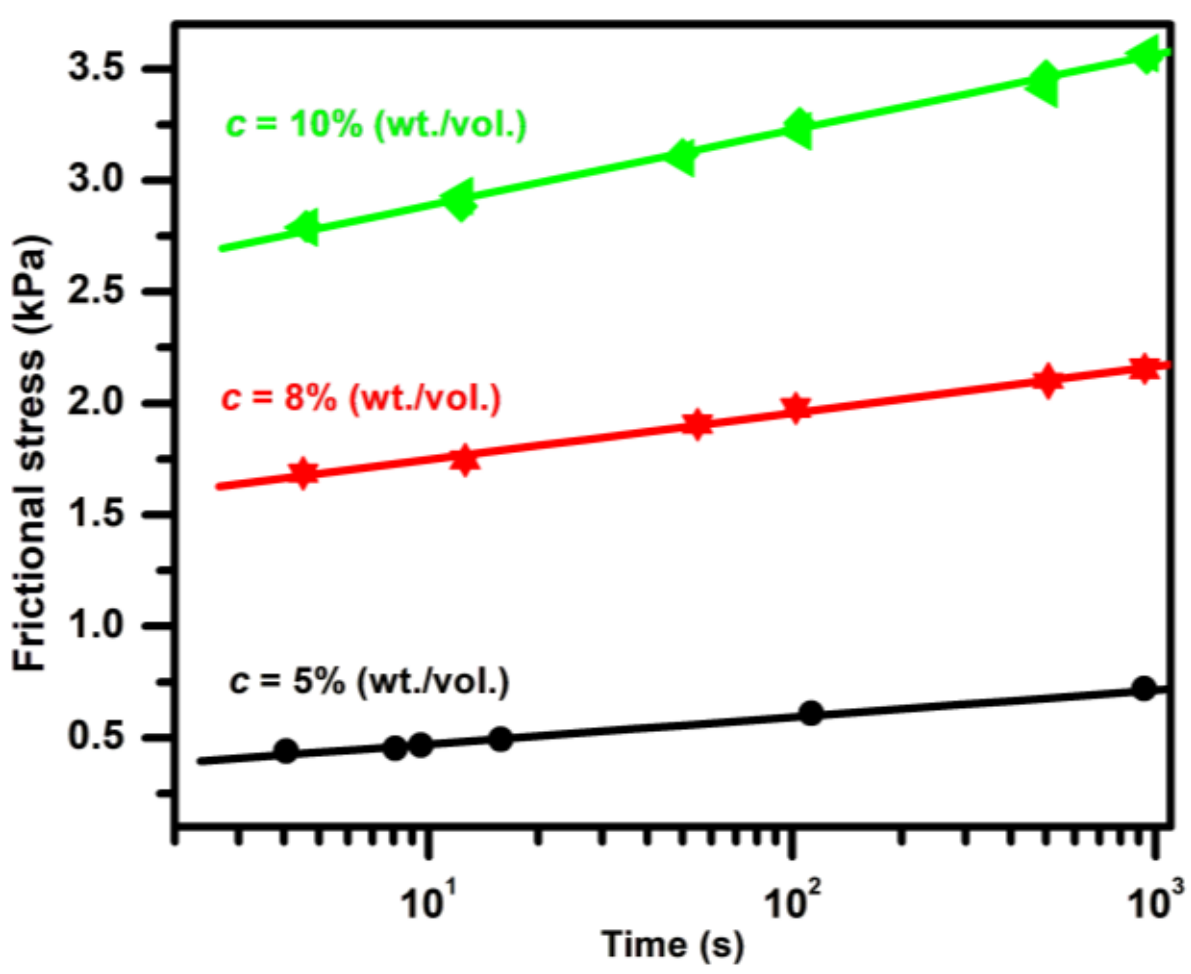

Fig.2. Frictional stress vs. a time for different aging time $t_w$ =5,12 & 72s of different gel concentration c=5%, 8% & 10% (wt./vol.). Experimental data from Baumberger et al. (2003).

Table 1 shows that $N_0$ increases with gelatin concentration in the hydrogels. This observation is attributed to increased number of dangling chains with increasing gelatin concentration (Singh et al., 2021). Moreover, ratio $u_0/\tau$ is found to be decreasing with increasing gelatin concentration in the hydrogel. This observation is attributed to increased relaxation time $\tau$ as shorter dangling chains find difficult to adhere with the substrate during the pulling process (Singh et al., 2021). While fast and slow aging constants $\tau_a$ and $\tau_b$ increase with gelatin concentration, since these constants are related to bonding process at $V_0$=0, that is, during stationary condition (Singh et al., 2021 & Singh and Juvekar, 2011).

Table 1 : Summary of the data obtained from the analysis of Fig.2.

| c %( wt./vol.) | 5% | 8% | 10% | |
|---|---|---|---|---|
| $N_0(\times 10^{12})$ | 1.46 | 4.76 | 7.27 | Increases |
| $u_0/\tau\,(s^{-1})$ | $2.90\times 10^{-6}$ | $0.796\times 10^{-6}$ | $0.060\times 10^{-6}$ | decreases |
| $\tau_a$(s) | 1.39 | 0.0194 | 0.0146 | decreases |
| $\tau_b$(s) | 1.80 | 0.0160 | 0.0095 | decreases |
| $G_d/h$(kP.m$^{-1}$) | 230 | 450 | 940 | Increases |
| Residual stress (kPa) | 0.343 | 1.54 | 2.57 | Increases |

It would be interesting to extend the present model on shear rate dependent static friction (Gupta and Singh, 2016). It will be really interesting to extend the present friction model on hard surfaces wherein microcontacts at sliding interface give rise to friction (Persson, 2013).

### 3.0 Conclusions:

Proposed friction model explains the experimental observation that peak of static friction increases with aging time at a fixed shear velocity. It is also established that static friction increases with gelatin concentrations. In future, it would be interesting to validate the present friction model as logarithm of shear rate at a fixed aging time.

### References

Berthoud, P., Baumberger, T., G'sell, C. and Hiver, J.M., 1999. Physical analysis of the state- and rate-dependent friction law: Static friction. *Physical review B*, *59*(22), p.14313.

Dieterich, J.H., 1978. Time-dependent friction and the mechanics of stick-slip. *Pure and applied geophysics*, *116*, pp.790-806.

Farain, K. and Bonn, D., 2023. Predicting frictional aging from bulk relaxation measurements. *Nature Communications*, *14*(1), p.3606.

Juvekar, V.A. and Singh, A.K., 2024. Analysis of Stick-Slip Motion as a Jump Phenomenon. arXiv preprint arXiv:2405.13142.

Li, Q., Tullis, T.E., Goldsby, D. and Carpick, R.W., 2011. Frictional ageing from interfacial bonding and the origins of rate and state friction. Nature, 480(7376), pp.233-236.

Ouyang, W. and Urbakh, M., 2022. Microscopic mechanisms of frictional aging. Journal of the Mechanics and Physics of Solids, 166, p.104944.

Petrova, D., Sharma, D.K., Vacha, M., Bonn, D., Brouwer, A.M. and Weber, B., 2020. Ageing of polymer frictional interfaces: the role of quantity and quality of contact. *ACS applied materials & interfaces*, *12*(8), pp.9890-9895.

Persson, B.N., 2013. Sliding friction: physical principles and applications. Springer Science & Business Media.

Ruina, A., 1983. Slip instability and state variable friction laws. *Journal of Geophysical Research: Solid Earth*, *88*(B12), pp.10359-10370.

Soni, P.K., Singh, A.K. and Katiyar, J.K., 2023. Experiments and prediction of hold time-dependent static friction of a wet granular layer. Tribology Letters, 71(3), p.75.

Singh, A.K. and Juvekar, V.A., 2011. Steady dynamic friction at elastomer–hard solid interface: a model based on population balance of bonds. Soft Matter, 7(22), pp.10601-10611.

Singh, A.K., Juvekar, V.A. and Bellare, J.R., 2021. A model for frictional stress relaxation at the interface between a soft and a hard solid. Journal of the Mechanics and Physics of Solids, 146, p.104191.